\documentclass[letterpaper, 10 pt, conference]{ieeeconf}  

\IEEEoverridecommandlockouts                              

\usepackage[intlimits]{amsmath}
\usepackage{amsfonts,amssymb}
\DeclareSymbolFontAlphabet{\mathbb}{AMSb}
\usepackage{float}
\usepackage[font=small,labelfont=bf]{caption}
\usepackage{subcaption}
\usepackage{fancyhdr}
\usepackage{fancybox}
\usepackage{ifthen}
\usepackage{lscape,afterpage}
\usepackage{xspace}
\usepackage{epstopdf}
\usepackage{cite}
\usepackage{textcomp}
\usepackage{tabularx}
\usepackage{adjustbox}
\usepackage{array}
\usepackage{makecell}
\usepackage{placeins}
\usepackage{pgf}
\usepackage{pgffor}
\usepackage{pgfmath}
\usepackage{algorithm}
\usepackage{algpseudocode}
\usepackage{lingmacros}
\usepackage{booktabs}

\usepackage{amsthm}
\usepackage{graphicx}
\usepackage[dvipsnames]{xcolor}
\usepackage{etoolbox}

\usepackage{xurl}
\usepackage{tikz}
\usepackage{multirow}
\usetikzlibrary{arrows.meta,positioning,calc,fit,backgrounds}
\makeatletter
\let\NAT@parse\undefined
\makeatother
\usepackage[hidelinks,hypertexnames=false]{hyperref}
\hypersetup{colorlinks=true, linkcolor=blue, citecolor=blue, urlcolor=blue}

\title{\LARGE \bf
Multi-Robot Multi-Queue Control via Exhaustive Assignment Actor-Critic Learning
\thanks{A version of this paper has been submitted to the 65th IEEE Conference on Decision and Control for possible publication.}
}

\author{Mohammad Merati$^{1}$, H. M. Sabbir Ahmad$^{1}$, Wenchao Li$^{2}$, and David Castañón$^{2}$
\thanks{$^{1}$Divisions of Systems Engineering, Boston University, 8 St Mary's St, Boston, MA 02215, United States. {\tt\small mmerati@bu.edu, sabbir92@bu.edu}}%
\thanks{$^{2}$Department of Electrical and Computer Engineering, Boston University, 8 St Mary's St, Boston, MA 02215, United States. {\tt\small wenchao@bu.edu,dac@bu.edu}}%
}

\begin{document}

\maketitle
\thispagestyle{empty}
\pagestyle{empty}

\begin{abstract}

We study online task allocation for multi-robot, multi-queue systems with asymmetric stochastic arrivals and switching delays. We formulate the problem in discrete time: each location can host at most one robot per slot, servicing a task consumes one slot, switching between locations incurs a one-slot travel delay, and arrivals at locations are independent Bernoulli processes with heterogeneous rates. Building on our previous structural result that optimal policies are of exhaustive type, we formulate a discounted-cost Markov decision process and develop an exhaustive-assignment actor-critic policy architecture that enforces exhaustive service by construction and learns only the next-queue allocation for idle robots. Unlike the exhaustive-serve-longest (ESL) queue rule, whose optimality is known only under symmetry, the proposed policy adapts to asymmetry in arrival rates. Across different server-location ratios, loads, and asymmetric arrival profiles, the proposed policy consistently achieves lower discounted holding cost and smaller mean queue length than the ESL baseline, while remaining near-optimal on instances where an optimal benchmark is available. These results show that structure-aware actor-critic methods provide an effective approach for real-time multi-robot scheduling.

\end{abstract}

\section{Introduction}

Real-time resource allocation under stochastic demand is a central challenge in many emerging engineered systems. In manufacturing, warehouse automation, and field robotics, mobile servers must repeatedly decide where to move and which workload to serve as tasks arrive over time in different spatial locations. These decisions must be made online with low latency, in response to arrival and completion of tasks. For such problems, exact optimization methods typically become computationally prohibitive in dynamic settings.  This difficulty is further amplified when task arrivals are spatially distributed and service resources are mobile, so that switching from one task location to another consumes time. Similar issues arise in the scheduling of multi-server, multi-queue systems where there are setup delays for servers to switch between queues.  

Dynamic vehicle routing problems with stochastic customer requests  are often posed as Markov Decision Processes  \cite{Ulmer2017DeliveryDeadlines}. However, the combinatorial complexity of deterministic vehicle routing problems and the curse of dimensionality leads researchers to the use of approximate dynamic programming (ADP) heuristics with significant online computation \cite{Ulmer2017DeliveryDeadlines,Ulmer2020HorizontalADP}.

When tasks arrive at fixed locations, one can formulate the resource allocation problem as scheduling servers in multi-class queuing networks, where each location represents a class of tasks. Finding optimal policies in multi-class queueing networks is difficult except in simple cases, and most models ignore the delays in switching servers to different classes.   
Channel allocation for mobile networks  \cite{lott2000optimality,Carroll2019ResponseDelays} is solved using index policies for bandit problems, where switching times associated with assigning channels are neglected.  For multiclass queueing problems, simple policies such as backpressure and max weight  \cite{shah2012,dai2005,Magluri2016} perform well when no switching delay is present.  For problems with switching delays, Hofri and Ross \cite{hofri_ross_1987} develop optimal control of a single server, two-queue system with server with switching times. This work was extended in \cite{liu1992optimal}, where optimal policies are computed for single server, multiple queue systems with exponential arrivals with the same arrival rate across queues ({\em homogeneous arrival rates}) and random switching delays.  In \cite{merati2025exhaustive}, these results are extended to multiple servers with homogeneous arrival rates and deterministic switching delays, resulting in an optimal policy ESL which is easily implemented.  In this paper, we are concerned with problems with multiple servers and queues, where the arrival rates are inhomogeneous and vary across queues. 

Due to the difficulty of obtaining optimal scheduling policies for more general multiclass queueing problems, several authors attempt to learn feedback policies using variations of reinforcement learning and dynamic programming.  Liu  \emph{et al.} \cite{liu2022rl} uses reinforcement learning to obtain policies for average-cost control of general queueing networks. Dai and Gluzman \cite{dai2022queueing} learn backlog-based server-scheduling policies from simulations using a deep actor-critic approach. Policy-gradient reinforcement learning has been used for adaptive packet routing in communication networks, where distributed routers learn stochastic forwarding policies from local observations and a global delay-based reward signal \cite{peshkin2002reinforcement}. \cite{staffolani2023rlq} uses reinforcement learning for workload allocation to distributed task queues with dedicated servers.  \cite{mitzenmacher2025queueing} provides an overview of recent results on learning scheduling approaches for multiclass queueing.   However, none of these approaches address problems with switching delays between classes, which is essential to capture the spatially distributed task arrivals that are of interest in this paper.

Recent studies have shown that exploiting characteristics of optimal policies in  learning approaches can result in simpler policy networks and much faster learning. In  \cite{comte2025score}, the authors exploit the structure of exponential-family stationary distributions to estimate gradients directly without the usual reinforcement-learning value-function approximations, and establish a simple learning policy with guaranteed optimality properties.  In \cite{jali2024efficient}, the authors learn routing policies for heterogeneous queueing systems with a single central queue, incorporating properties that the policy must be of soft-threshold type.  In \cite{wigmore2025novel}, the author embeds switch-type monotonicity directly into the policy network, and show that the resulting policies have improved sample efficiency and generalization over standard multi-layer perceptron architectures.  Although these results do not address queues with switching delays, they suggest approaches for integrating policy structure into learning architectures.  

In this paper, we formulate the asymmetric multi-robot multi-queue scheduling problem with stochastic arrivals and switching delays as a discounted Markov decision process, as in our previous work \cite{merati2025exhaustive}.  We established in \cite{merati2025exhaustive} that optimal policies must be of exhaustive type, where a server does not switch queues until the queue it is currently serving is exhausted.  
We propose an exhaustive-assignment actor-critic (EA-AC) architecture, trained using Proximal Policy Optimization (PPO), which integrates this policy structure for the present problem. The actor enforces exhaustive service by construction and performs sequential masked assignment of idle robots to feasible queues, while the critic uses a centralized state encoder to estimate the global cost-to-go. 
We compare the performance of our learned policy with optimal policies computed by stochastic dynamic programming on small examples, and find that the learned policy performs as well as the optimal policy in these examples.  We also compare the performance of the learned policy to that of the ESL policy of \cite{merati2025exhaustive} in larger examples with homogeneous arrival processes across queues, and find the performance of the two policies is statistically indistinguishable.  Finally, we compare the performance of the learned policy and the ESL policy on large examples with inhomogenous arrival process.  As expected, the learned policy significantly outperforms the ESL policy, as the ESL policy is not optimal for these cases.  These results establish that our learned policy approach provides a viable design for feedback control of multi-server, multi-queue systems with inhomogenous task arrival processes and switching delays.  

The rest of the paper is organized as follows: Section~\ref{sec:model} presents the mathematical decision problem. Section~\ref{sec:method} presents the learning-based policy design, including the PPO-based actor-critic formulation, the exhaustive assignment policy network, and the training procedure. Section~\ref{sec:experiments} reports simulation results and compares our proposed algorithm against baseline policies.  Section~\ref{sec:conclusion} has our conclusions and directions for future work.
\section{Problem Formulation}\label{sec:model}

In this section, we describe a discrete-time Markov decision process (MDP) for a team of mobile robots serving spatially distributed task queues.\footnote{We use the terms robots/servers and queues/locations interchangeably.} We consider a system with $M$ robots and $N$ locations, under the constraint that no more than one robot can occupy a location in any slot. The objective is to minimize a discounted holding cost.

\subsection{System description}
The system operates in discrete time. There are $N$ locations $\{\mathcal{Q}_1,\dots,\mathcal{Q}_N\}$ and $M$ \emph{non-preemptive} robots, where $M\le N$. A robot can complete at most one task in a time slot. 

Let $x_i(t)\in\mathbb{N}_0$ denote the number of waiting tasks at location $i$ at the beginning of time $t$, and let $s_r(t)\in\{1,\dots,N\}$ denote the location of robot $r$ at the same time. The system state is therefore written as $z(t)=\bigl(s(t);x(t)\bigr)$, where $s(t)=(s_1(t),\dots,s_M(t))$ and $x(t)=(x_1(t),\dots,x_N(t))$.

Tasks arrive independently across locations according to Bernoulli processes. At each time $t$, an arrival  $a_i(t)\sim\text{Bernoulli}(p_i),~i=1,\dots,N$ occurs, where $a_i(t)=1$ when a new task arrives to location $i$ in time $t$, and $a_i(t)=0$ indicates that no new task arrives. We use the \emph{late-arrival convention}: if a task arrives during time $t$, it joins the queue at the \emph{end} of that slot and cannot be served until slot $t+1$.

Service times are deterministic and equal to one time period. Thus, if a robot starts serving at the beginning of time $t$ at its current location, it completes exactly one task there by the end of the slot. If instead the robot switches to another location, then it incurs a \emph{deterministic one-slot travel delay}, during which no service is provided. In addition, at most one robot can be assigned to any location at each time.

\subsection{State dynamics}

At the \emph{start} of time $t$, the state of the system is
\begin{equation*}
z(t)=\Bigl(\,s_1(t),\dots,s_M(t)\,;\;x_1(t),\dots,x_N(t)\Bigr)
\end{equation*}

At the beginning of slot $t$, a control action $u(t)=\bigl(u_1(t),\dots,u_M(t)\bigr)$ is selected. For each robot $r$, the admissible action set is
\begingroup\small
\begin{align*}
u_r(t)\in
\begin{cases}
\{\textbf{serve},\textbf{idle},\textbf{switch}(j): j\neq s_r(t)\} & x_{s_r(t)}(t)>0\\
\{\textbf{idle},\textbf{switch}(j): j\neq s_r(t)\} & x_{s_r(t)}(t)=0
\end{cases}
\end{align*}
\endgroup
subject to the system constraint that at most one robot may occupy a location in any slot. If robot $r$ is currently at location $i$ and takes action $u_r(t)=\textbf{serve}$, then one task is completed from queue $i$ during slot $t$. If $u_r(t)=\textbf{switch}(j)$, then robot $r$ spends the entire slot traveling to location $j$. If $u_r(t)=\textbf{idle}$, no service takes place during that slot.

Motivated by the structural results in \cite{merati2025exhaustive}, the learning-based policy developed in this paper restricts attention to \emph{exhaustive} policies. Under this restriction, a robot located at a nonempty queue continues serving that queue until it becomes empty, so the effective decision set becomes
\begin{align*}
u_r(t)\in
\begin{cases}
\{\textbf{serve}\} & x_{s_r(t)}(t)>0\\
\{\textbf{idle},\textbf{switch}(j): j\neq s_r(t)\} & x_{s_r(t)}(t)=0
\end{cases}
\end{align*}
Thus, the proposed method does not search over all admissible controls of the MDP, but rather over the smaller class of exhaustive policies, which substantially reduces the assignment decision space.

Given the control actions and feasibility conditions above, the amount of service completed at location $i$ during slot $t$ is summarized by the departure indicator
\[
  d_i(t)
  := \sum_{r=1}^M \mathbf 1\{\,s_r(t)=i,\ u_r(t)=\textbf{serve}\,\}
  \in \{0,1\}
\]
The queue-length and robot-location dynamics are then described by the recursion
\begingroup\small
\begin{equation}\label{eq:MsrvNq-update-nocoloc}
\begin{aligned}
  x_i(t+1) &= x_i(t) - d_i(t) + a_i(t) && i=1,\dots,N\\[4pt]
  s_r(t+1) &=
  \begin{cases}
     s_r(t) & u_r(t)\in\{\textbf{serve},\textbf{idle}\}\\[2pt]
     j      & u_r(t)=\textbf{switch}(j)
  \end{cases}
  && r=1,\dots,M
\end{aligned}
\end{equation}
\endgroup

\subsection{Control Objective}
Let $\mathcal S=\{1,\dots,N\}^{M}\times \{0,1,2,\dots\}^{\,N}$
denote the state space, and let $\mathcal U(z)$ denote the set of admissible controls at state $z$. For a stationary deterministic policy $\pi:\mathcal S\to\mathcal U(z)$, define the one-step holding cost by $c\bigl(z(t)\bigr)=\sum_{i=1}^{N} x_i(t)$. The discounted cost-to-go under policy $\pi$ from initial state $z$ is
\[
V_\pi(z)=\mathbb E_z^{\pi}\Biggl[\sum_{t=0}^{\infty} \beta^{t} c\bigl(z(t)\bigr)\Biggr]
\]
where $\beta\in(0,1)$ is the discount factor. The control objective is to find a policy that minimizes this expected discounted holding cost, namely
\[
V^*(z)=\inf_{\pi} V_\pi(z).
\]

\section{Learning-Based Policy Design}\label{sec:method}

In this section, we describe the proposed EA-AC learning framework. We first present the PPO-based actor-critic formulation, then describe the exhaustive-assignment policy network, and finally summarize the training procedure.

\subsection{Actor-Critic Policy Optimization via PPO}

We model the asymmetric multi-robot multi-queue control problem as a discounted Markov decision process and parameterize the control law by a stochastic policy $\pi_\theta(a\mid z)$, where $z$ denotes the system state and $a$ denotes a feasible joint action. To optimize this policy, we adopt PPO \cite{schulman2017proximal}, a first-order policy gradient method within the actor-critic family that jointly learns a stochastic policy (actor) $\pi_\theta$ and a value function (critic) $V_\phi(z)$. PPO alternates between sampling trajectories $\{(z_t,a_t,r_t)\}$ under the current policy and performing stochastic gradient updates using mini-batches of this data. The critic is trained to estimate $V_\phi(z)$, which is an approximation of the discounted value function $$V^\pi(z)=\mathbb{E}_\pi[\sum_{l=0}^{\infty}\beta^l r_{t+l}\mid z_t=z]$$
which is used as a baseline for variance reduction. In practice, PPO estimates the advantage from rollout data using generalized advantage estimation (GAE), which can be written in the form $\hat A_t = \hat R_t - V_\phi(z_t)$,
where $\hat R_t$ denotes the corresponding bootstrapped return target. Intuitively, $\hat A_t$ measures whether action $a_t$ performs better or worse than the average behavior at state $z_t$.

The policy is updated by maximizing a clipped surrogate objective that constrains the deviation from the behavior policy used to generate the data. Defining the likelihood ratio $r_t(\theta)=\frac{\pi_\theta(a_t\mid z_t)}{\pi_{\theta_{\mathrm{old}}}(a_t\mid z_t)}$, PPO maximizes
\begin{align*}
L^{\mathrm{clip}}(\theta)=
\mathbb{E}\Big[
\min\big(
r_t(\theta)\hat A_t,\;
\mathrm{clip}(r_t(\theta),1-\epsilon,1+\epsilon)\hat A_t
\big)
\Big]
\end{align*}
where the expectation is taken over samples generated by $\pi_{\theta_{\mathrm{old}}}$. The clipping operation limits large policy updates by truncating the incentive to increase or decrease action probabilities beyond a prescribed range, to stabilize training while retaining the simplicity of first-order optimization.

\begin{figure}
    \centering
    \includegraphics[width=1\linewidth]{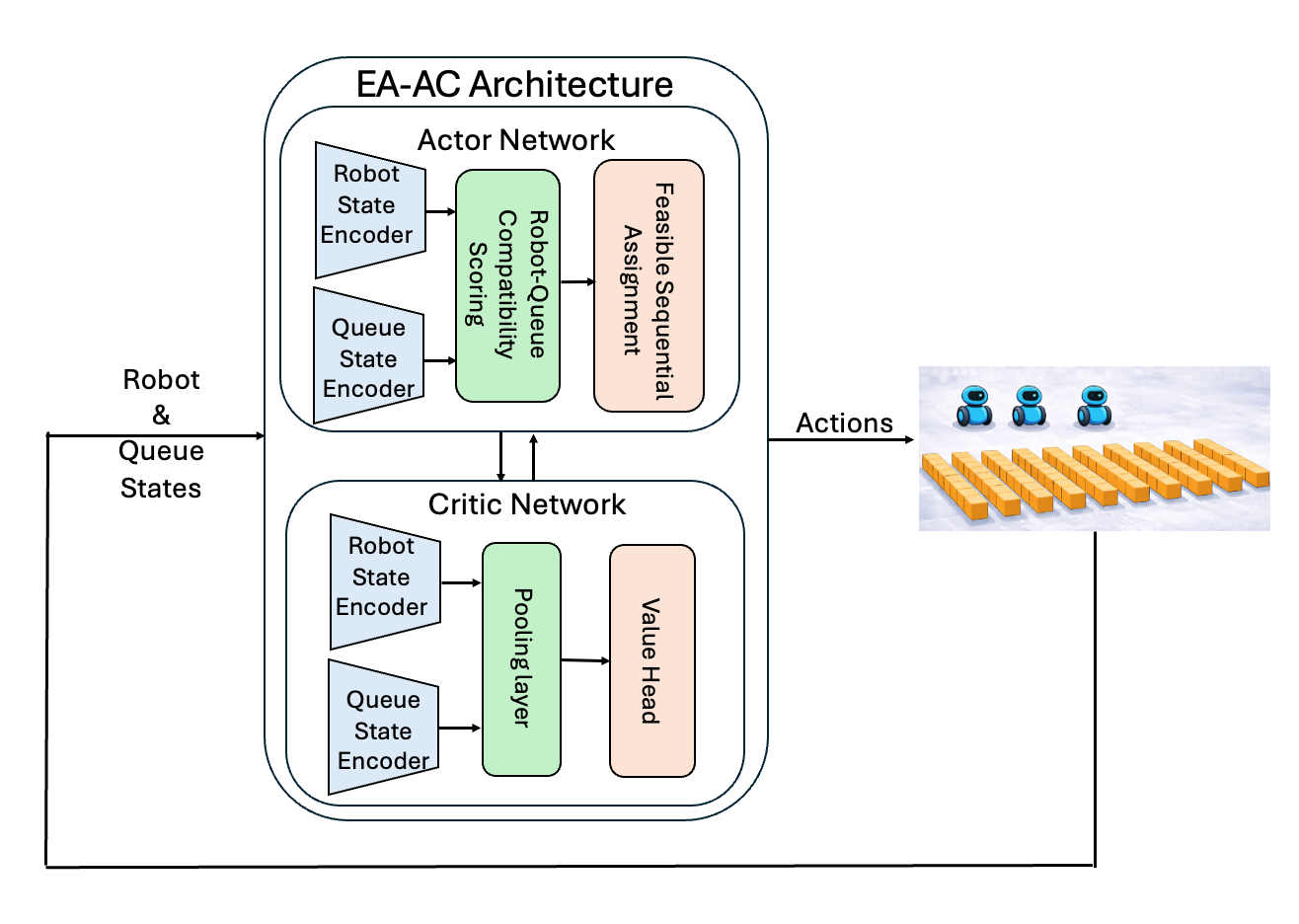}
    \caption{Block diagram of the proposed EA-AC architecture. The policy network encodes robot and queue states, models robot-queue interactions, and generates sequential assignment decisions for idle robots. The critic network uses centralized state information to estimate the value function during training.}
    \label{fig:eaac_architecture}
\end{figure}

In our setting, the actor maps the state to a distribution over feasible idle-robot assignment decisions under the exhaustive-service restriction, while the critic estimates the corresponding discounted holding cost-to-go. The actor therefore learns how to reallocate idle robots across locations to reduce future congestion, whereas the critic evaluates the long-term effect of these decisions under the discounted objective. Since the decision space is discrete and combinatorial, the policy acts by assigning probabilities over feasible assignments rather than producing continuous controls. PPO is a natural choice in this setting because it enables stable first-order policy optimization from simulated trajectories and is readily compatible with the structured policy parameterization developed in the next subsection.

\subsection{Exhaustive Assignment Policy Network}\label{sec:policy_network}

Our policy architecture is built around the structural restriction that service is exhaustive: once a robot is located at a nonempty queue, it continues serving that queue until the queue becomes empty. Under this restriction, the control problem is reduced from choosing a joint action for all robots at every slot to choosing new destinations only for robots whose current queues are empty. Thus, rather than learning an unconstrained joint control law over all robots, the actor only needs to learn how to reassign idle robots to feasible destination queues.

Let
\[
\mathcal{B}(z)=\{r: x_{s_r}>0\}, \qquad \mathcal{I}(z)=\{r: x_{s_r}=0\}
\]
denote the sets of \emph{busy} and \emph{idle} robots in state $z=(s_1,\dots,s_M;x_1,\dots,x_N)$. For each $r\in\mathcal{B}(z)$, the action is fixed by the exhaustive-service rule, and only robots in $\mathcal{I}(z)$ require an assignment decision. Accordingly, the policy is parameterized in the factorized form
\begin{align}
\pi_\theta(u\mid z)
&=
\Biggl(\prod_{r\in\mathcal{B}(z)} \mathbf{1}\{u_r=s_r\}\Biggr)
\notag\\
&\quad\times
\Biggl(\prod_{k=1}^{|\mathcal{I}(z)|}
\pi_\theta\!\left(u_{\rho_k}\mid z,u_{\rho_1},\dots,u_{\rho_{k-1}}\right)\Biggr)
\end{align}
where $\rho_1,\dots,\rho_{|\mathcal{I}(z)|}$ is a fixed ordering of the idle robots. This factorization removes forced decisions from the learning problem and concentrates the policy representation on the only nontrivial part of the control law, namely the reassignment of idle robots to new queues.

The actor uses two sets of learned embeddings, one for queues and one for robots. For each queue $i$, we form a queue feature vector
\[
\xi_i = \bigl[\bar x_i,\ \bar\lambda_i,\ o_i,\ 1-o_i\bigr]
\]
where $\bar x_i$ is the queue length normalized by the queue maximum length, $\bar\lambda_i$ is the arrival rate normalized by the maximum arrival rate in the system, and $o_i\in\{0,1\}$ indicates whether queue $i$ is currently occupied by a robot. These features encode the local congestion level, arrival intensity, and current feasibility status of queue $i$. The queue features are mapped through a shared multilayer perceptron to obtain queue tokens
\[
h_i = \phi_q(\xi_i)\in\mathbb{R}^d, \qquad i=1,\dots,N.
\]

For each robot $r$, we define a robot feature vector
\[
\eta_r = \bigl[e(s_r),\ \bar x_{s_r},\ \bar\lambda_{s_r},\ b_r\bigr]
\]
where $e(s_r)$ is a learned embedding of the robot's current location, $\bar x_{s_r}$ and $\bar\lambda_{s_r}$ are the normalized queue length and arrival rate at the location of robot $r$, and $b_r=\mathbf{1}\{x_{s_r}>0\}$ is the busy indicator. These features summarize the robot's current context: where it is, what queue condition it currently faces, and whether its action is fixed by exhaustive service. They are passed through a second shared multilayer perceptron to obtain robot tokens
\[
g_r = \phi_r(\eta_r)\in\mathbb{R}^d, \qquad r=1,\dots,M.
\]

For an idle robot $r$ and candidate destination $i$, the actor computes a compatibility score
\[
\ell_{r,i}
=
\frac{\langle g_r,h_i\rangle}{\sqrt{d}} + c_i
\]
where $c_i$ is a learned scalar bias associated with queue $i$. These scores define the logits of the assignment distribution and quantify how appropriate it is to send robot $r$ to queue $i$ given both the robot state and the queue state.

Two levels of masking are then applied. First, a \emph{current-occupancy mask} removes queues that are already occupied at the current slot, except that staying at the robot's current location remains feasible. This mask enforces the one-robot-per-queue constraint at the level of the policy output. Second, the sequential decoder applies a \emph{reservation mask}: once an idle robot is assigned to a queue, that queue is removed from the feasible action set of the remaining idle robots. This guarantees that no two idle robots are assigned to the same destination within the same decision step. For robots in $\mathcal{B}(z)$, the actor does not produce a free assignment; instead, their logits are replaced by a degenerate distribution concentrated on the current queue, thereby enforcing exhaustive service exactly within the policy parameterization.

The critic is centralized and estimates the discounted value of the full system state. It uses separate queue and robot encoders, but unlike the actor it does not construct assignments. The reason is that the critic is not required to choose a feasible reassignment; rather, it must evaluate the long-term congestion consequences of the overall state. Queue tokens are built from queue-level features and robot tokens from robot-level features, and these are pooled across all queues and all robots to obtain permutation-invariant summaries of the state. In addition to these pooled embeddings, we include a small set of global backlog statistics,
\[
g(z)=\Bigl[\sum_{i=1}^N \bar x_i,\ \max_{1\le i\le N}\bar x_i,\ \frac{1}{N}\sum_{i=1}^N \bar x_i,\ \frac{|\mathcal{I}(z)|}{M}\Bigr]
\]
which summarize the total congestion level, the largest queue, the mean queue length, and the fraction of idle robots. The critic output is then given by
\[
V_\phi(z)=\psi\!\left(\bar h^{\,q},\bar h^{\,r},g(z)\right),
\]
where $\bar h^{\,q}$ and $\bar h^{\,r}$ denote the pooled queue and robot embeddings and $\psi$ is a feedforward value head. This centralized value estimator is used only to provide advantage estimates for the PPO update, while the actor remains responsible for the structured reassignment decisions described above. Additional implementation details of the actor and critic networks are reported in the Appendix~\ref{app:network}.

\begin{table*}[t]
\centering
\caption{Performance comparison between ESL and EA-AC across asymmetric scenarios. Values are reported as mean $\pm$ 95\% CI half-width. Best values in each scenario are shown in bold. Improvement is computed relative to ESL.}
\label{tab:main_results}
\small
\setlength{\tabcolsep}{6pt}
\renewcommand{\arraystretch}{1.2}
\begin{tabular}{ccccccc}
\hline
Scenario & $(M,N)$ & Policy & Discounted Cost ($\downarrow$) & Mean Queue Length ($\downarrow$) & Cost Red. (\%) ($\uparrow$) & Queue Red. (\%) ($\uparrow$) \\
\hline

\multirow[c]{2}{*}{S1} & \multirow[c]{2}{*}{$(1,3)$}
& ESL & $410.51 \pm 8.67$ & $1.6133 \pm 0.0214$ & -- & -- \\
& & EA-AC & $\mathbf{405.99 \pm 8.64}$ & $\mathbf{1.5905 \pm 0.0211}$ & $\mathbf{1.10}$ & $\mathbf{1.41}$ \\

\hline

\multirow[c]{2}{*}{S2} & \multirow[c]{2}{*}{$(6,24)$}
& ESL & $1521.64 \pm 14.44$ & $0.6927 \pm 0.0035$ & -- & -- \\
& & EA-AC & $\mathbf{1432.86 \pm 13.80}$ & $\mathbf{0.6381 \pm 0.0033}$ & $\mathbf{5.83}$ & $\mathbf{7.88}$ \\

\hline

\multirow[c]{2}{*}{S3} & \multirow[c]{2}{*}{$(12,60)$}
& ESL & $3757.98 \pm 24.87$ & $0.7047 \pm 0.0024$ & -- & -- \\
& & EA-AC & $\mathbf{3451.51 \pm 26.29}$ & $\mathbf{0.6362 \pm 0.0026}$ & $\mathbf{8.16}$ & $\mathbf{9.71}$ \\

\hline

\multirow[c]{2}{*}{S4} & \multirow[c]{2}{*}{$(20,120)$}
& ESL & $7171.24 \pm 36.37$ & $0.6861 \pm 0.0018$ & -- & -- \\
& & EA-AC & $\mathbf{6522.10 \pm 40.68}$ & $\mathbf{0.6178 \pm 0.0021}$ & $\mathbf{9.05}$ & $\mathbf{9.96}$ \\

\hline

\multirow[c]{2}{*}{S5} & \multirow[c]{2}{*}{$(35,140)$}
& ESL & $10615.24 \pm 44.17$ & $0.8459 \pm 0.0019$ & -- & -- \\
& & EA-AC & $\mathbf{10060.60 \pm 43.59}$ & $\mathbf{0.7970 \pm 0.0018}$ & $\mathbf{5.23}$ & $\mathbf{5.78}$ \\

\hline

\multirow[c]{2}{*}{S6} & \multirow[c]{2}{*}{$(50,200)$}
& ESL & $12241.40 \pm 37.60$ & $0.6726 \pm 0.0008$ & -- & -- \\
& & EA-AC & $\mathbf{10745.19 \pm 42.92}$ & $\mathbf{0.5742 \pm 0.0011}$ & $\mathbf{12.22}$ & $\mathbf{14.63}$ \\

\hline

\multirow[c]{2}{*}{S7} & \multirow[c]{2}{*}{$(75,350)$}
& ESL & $21147.36 \pm 49.18$ & $0.6714 \pm 0.0006$ & -- & -- \\
& & EA-AC & $\mathbf{19287.91 \pm 56.68}$ & $\mathbf{0.6057 \pm 0.0009}$ & $\mathbf{8.89}$ & $\mathbf{9.92}$ \\

\hline
\end{tabular}
\end{table*}

The resulting architecture combines three elements that are specific to the present control problem: (i) exhaustive service is enforced directly at the policy level, (ii) switching decisions are represented as a sequential assignment of idle robots to distinct feasible queues, and (iii) the critic evaluates the global congestion state through pooled state summaries rather than through an unconstrained joint-action representation. In this way, the policy class is restricted to controls that respect the operational structure of the system while remaining flexible enough to learn reassignment rules that improve upon greedy switching in asymmetric environments.

\subsection{Training Procedure}\label{sec:training_procedure}

For each environment configuration, defined by the number of robots $M$, the number of locations $N$, and the arrival-rate vector $p=(p_1,\dots,p_N)$, we train a separate actor-critic policy using the architecture described above. In all training runs, the system starts from the empty state, the discount factor is fixed at $\beta=0.99$, and the immediate reward at time $t$ is taken to be the negative one-step holding cost $r_t = -\sum_{i=1}^N x_i(t)$, and no additional reward shaping is used.

In simulation, the infinite-capacity, infinite-horizon model of Section~\ref{sec:model} is approximated by a finite-capacity, finite-horizon implementation with queue cap $Q_{\max}=100$ and episode length $T=1000$. These values are chosen large enough that queue saturation is negligible and truncation effects do not materially affect the reported results.

The policy and value networks are optimized jointly using Adam with learning rate $7\times 10^{-4}$. The PPO clipping parameter is set to $\epsilon=0.2$, the value-loss coefficient to $0.5$, the entropy coefficient to $10^{-3}$, and the gradient norm is clipped at $0.5$. For evaluation, we use the final trained policy in deterministic mode: busy robots remain at their current queues by construction, and idle robots are assigned to the highest-probability feasible destinations under the sequential masked decoder.

\section{Experiments and Results}\label{sec:experiments}

In this section, we evaluate the proposed EA-AC policy on asymmetric multi-robot multi-queue systems and compare it against the ESL baseline. We compare EA-AC with optimal benchmarks available in small asymmetric instances through dynamic programming and in symmetric instances where ESL is known to be optimal. These experiments assess both policy quality and scalability as the system size increases.

\subsection{Simulation Scenarios and Baselines}

\begin{figure*}[t]
    \centering
    \includegraphics[width=\textwidth]{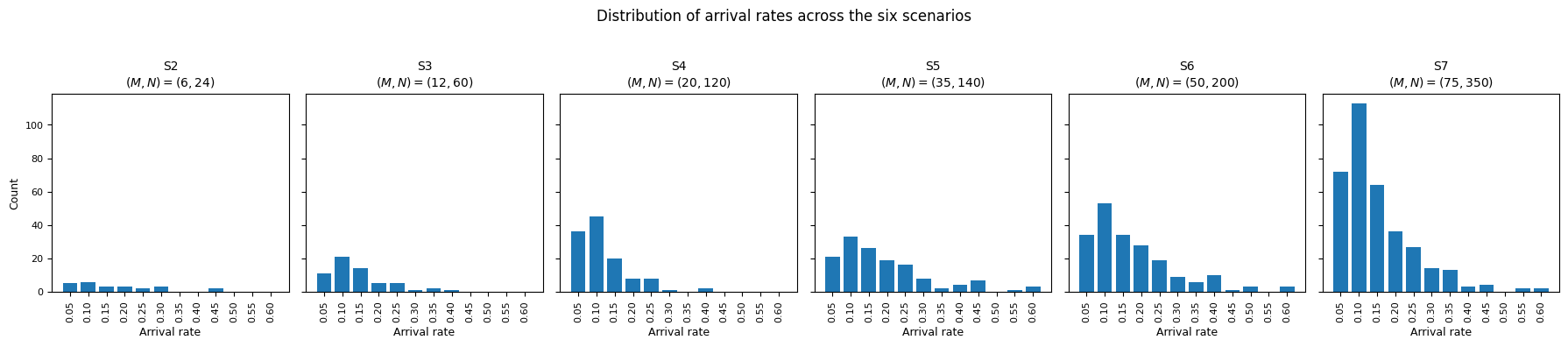}
    \caption{Distribution of arrival rates across the six asymmetric scenarios. Each bar indicates the number of queues whose arrival rate falls at a given value on the $0.05$ grid.}
    \label{fig:arrival_profiles}
\end{figure*}

We consider multiple asymmetric system configurations with different numbers of robots and queues. For each experiment, the arrival-rate vector $\boldsymbol{\lambda}=(\lambda_1,\dots,\lambda_N)$ is generated subject to a fixed total offered load and per-queue bounds. Specifically, we enforce $\sum_{i=1}^N \lambda_i = M\rho$,
where $M$ is the number of robots and $\rho$ is the prescribed load parameter, together with
\[
\lambda_{\min}\le \lambda_i \le \min(\lambda_{\max},\rho), \qquad
\lambda_{\min}=0.05,\;\lambda_{\max}=0.6.
\]
To introduce heterogeneity across queues, we first sample a weight vector from a symmetric Dirichlet distribution, scale it to the target total load, and then quantize the resulting rates to the $0.05$ grid while preserving the load and bound constraints. Across the reported scenarios, the load parameter lies in the range $[0.7,0.8]$. Figure~\ref{fig:arrival_profiles} shows the resulting distribution of arrival rates for each scenario. The same policy architecture is used across all problem sizes, allowing us to assess how the method scales with the numbers of robots and queues.

We compare the proposed EA-AC policy against the ESL baseline. Under ESL, each robot serves its current queue whenever that queue is nonempty; otherwise, an idle robot is assigned to the currently unoccupied nonempty queue with the largest queue length, with ties broken by higher arrival rate and then by queue index. For the proposed policy, the actor-critic network is trained separately for each scenario using the procedure of Section~\ref{sec:training_procedure}, and the learned policy is evaluated deterministically at test time. For small instances where an optimal policy is computable, we also compare EA-AC against the corresponding optimal benchmark.

\subsection{Evaluation Metrics}

Our primary performance metrics are the mean queue length and the total discounted holding cost. For a simulated trajectory of length $T$, the mean queue length is defined as
\[
\bar q = \frac{1}{T}\sum_{t=0}^{T-1}\frac{1}{N}\sum_{i=1}^N x_i(t)
\]
which measures the average congestion level in the system. The total discounted cost is defined as
\[
V = \sum_{t=0}^{T-1}\beta^t \sum_{i=1}^N x_i(t)
\]
with discount factor $\beta=0.99$, and is the main objective used for policy optimization and comparison.

For each scenario and each policy, results are averaged over 500 independent simulation runs, and 95\% confidence intervals are reported. We also monitor queue-cap occupancy to confirm that the reported performance differences are not driven by artificial queue saturation.

\subsection{Results}

Table~\ref{tab:main_results} summarizes the performance of the proposed EA-AC policy and the ESL baseline across the scenarios. In every case, EA-AC achieves both lower discounted holding cost and lower mean queue length than ESL. The improvement is modest in the smallest system, S1, but becomes much more pronounced in the larger scenarios. The largest gain is observed in S6, where EA-AC reduces discounted cost by $12.22\%$ and mean queue length by $14.63\%$ relative to ESL. Substantial improvements are also obtained in S4, with reductions of $9.05\%$ in discounted cost and $9.96\%$ in mean queue length. Even in the largest configuration, $(M,N)=(75,350)$, EA-AC continues to improve upon ESL by $8.89\%$ in discounted cost and $9.92\%$ in mean queue length.

The variation in performance gain across scenarios appears to be related not only to problem size, but also to the structure of the arrival-rate distribution. In scenarios such as S4 and S6, the arrival profiles exhibit stronger heterogeneity, so assigning idle robots using only the longest currently available queue is less effective than using a learned reassignment rule that can account for both backlog and long-run arrival asymmetry. By contrast, the gain in S5 is smaller, despite the larger system size. A plausible explanation is that the arrival-rate distribution in S5 is more diffuse across moderate values and hence offers less clear separation between highly critical and less critical queues. Also, S5 operates at a relatively congested regime, as reflected by the larger mean queue lengths under both policies, which likely reduces the frequency of idle-robot reassignment decisions and therefore narrows the margin over ESL. These observations suggest that the benefit of EA-AC is largest when asymmetry is both pronounced and operationally exploitable through reassignment.

Overall, the results show that the proposed method consistently improves upon the ESL rule in asymmetric systems. This provides empirical evidence that, although exhaustive service remains a useful structural principle, selecting the longest available queue is generally suboptimal once symmetry is lost. The results also demonstrate a favorable scalability trend: the same compact neural-network architecture is used in all scenarios, yet it remains effective from very small to fairly large systems. Thus, the observed gains are achieved not by redesigning the model for each problem scale, but by using a single structured policy class that adapts to different system sizes and asymmetric arrival-rate profiles. Additional performance comparisons across a broader set of robot--queue configurations are provided in Appendix~\ref{app:additional_results}.

Figure~\ref{fig:convergence_scaling} shows the empirical number of training iterations required for convergence as the system size increases. Across the tested scenarios, this quantity grows approximately linearly with the number of robots, while variation in the queue-to-robot ratio within the tested range has a comparatively smaller effect. This suggests that, for the proposed EA-AC architecture, convergence behavior is driven primarily by the number of active decision-makers. Although we do not claim a formal sample-complexity guarantee, the observed trend provides evidence that the method scales well in practice.

\begin{figure}
    \centering
    \includegraphics[width=1\linewidth]{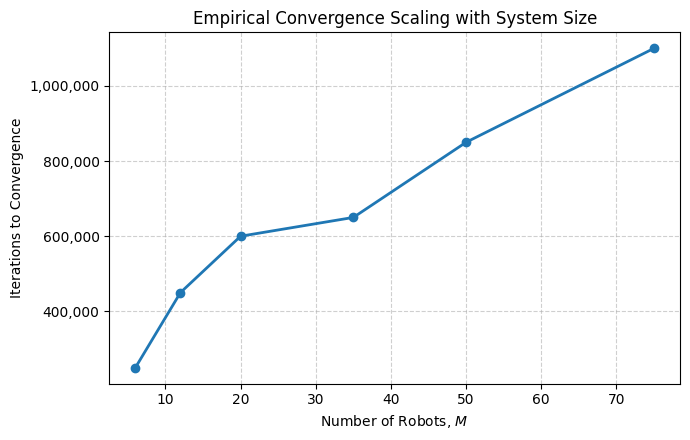}
    \caption{Empirical convergence scaling of the proposed EA-AC policy with respect to the number of robots. Over the tested scenarios, the number of training iterations required for convergence grows approximately linearly with system size.}
    \label{fig:convergence_scaling}
\end{figure}

\begin{table*}[t]
\centering
\caption{Comparison of the proposed EA-AC policy with optimal benchmarks. For S1--S3, the reference policy is the exact optimal policy; for S4--S5, it is the ESL policy, which is optimal under symmetry.}
\label{tab:small_optimal}
\small
\setlength{\tabcolsep}{6pt}
\renewcommand{\arraystretch}{1.2}
\begin{tabular}{c c c c c c}
\hline
Scenario & $(M,N)$ & Arrival rates $p$ & Policy & Discounted Cost ($\downarrow$) & Mean Queue Length ($\downarrow$) \\
\hline

\multirow{2}{*}{S1} & \multirow{2}{*}{$(1,3)$} & \multirow{2}{*}{$[0.10,\ 0.25,\ 0.45]$}
& Optimal & $\mathbf{405.8574 \pm 8.68}$ & $\mathbf{1.5895 \pm 0.0211}$ \\
& & & EA-AC & $405.9872 \pm 8.64$ & $1.5905 \pm 0.0211$ \\

\hline

\multirow{2}{*}{S2} & \multirow{2}{*}{$(1,4)$} & \multirow{2}{*}{$[0.05,\ 0.10,\ 0.25,\ 0.30]$}
& Optimal & $\mathbf{322.3632 \pm 6.27}$ & $\mathbf{0.8875 \pm 0.0102}$ \\
& & & EA-AC & $322.5075 \pm 6.27$ & $0.8877 \pm 0.0101$ \\

\hline

\multirow{2}{*}{S3} & \multirow{2}{*}{$(2,4)$} & \multirow{2}{*}{$[0.15,\ 0.25,\ 0.50,\ 0.60]$}
& Optimal & $\mathbf{397.0253 \pm 5.15}$ & $\mathbf{1.0320 \pm 0.0065}$ \\
& & & EA-AC & $399.4329 \pm 5.12$ & $1.0380 \pm 0.0066$ \\

\hline

\multirow{2}{*}{S4} & \multirow{2}{*}{$(6,36)$} & \multirow{2}{*}{$[0.1167]^{36}$}
& ESL & $2350.8514 \pm 23.98$ & $0.7529 \pm 0.0044$ \\
& & & EA-AC & $\mathbf{2349.5496 \pm 23.89}$ & $\mathbf{0.7529 \pm 0.0044}$ \\

\hline

\multirow{2}{*}{S5} & \multirow{2}{*}{$(12,60)$} & \multirow{2}{*}{$[0.14]^{60}$}
& ESL & $\mathbf{3961.0604 \pm 29.06}$ & $\mathbf{0.7368 \pm 0.0027}$ \\
& & & EA-AC & $3962.1129 \pm 28.85$ & $0.7369 \pm 0.0027$ \\

\hline
\end{tabular}
\end{table*}

\subsection{Comparison with Optimal Benchmarks}

To assess the quality of the proposed policy, we also compare it against optimal references on instances where such a benchmark is available. Table~\ref{tab:small_optimal} reports the discounted cost and mean queue length of the proposed policy on five scenarios. For S1--S3, the reference policy is the exact optimal policy computed on small instances. For S4--S5, the instances satisfy the symmetry conditions under which the ESL policy is known to be optimal, so ESL serves as the corresponding optimal benchmark.

Across all five scenarios, the learned policy remains extremely close to the optimal reference. In the single-robot cases S1 and S2, the gap between EA-AC and the optimal policy is negligible on both discounted cost and mean queue length. In the two-robot asymmetric case S3, the learned policy still remains very close to optimal, with only a small increase in both metrics. The same behavior is observed in the larger symmetric cases S4 and S5. There, the EA-AC policy is nearly indistinguishable from ESL, with differences that are negligible relative to the reported confidence intervals and therefore not statistically meaningful. In particular, the two policies produce essentially identical mean queue lengths, while the discounted-cost differences are on the order of the estimation noise.

Taken together, these results indicate that the proposed EA-AC architecture recovers near-optimal behavior whenever an optimal benchmark is available. This supports its use as a practical approximation method in larger asymmetric systems, where exact dynamic programming is computationally intractable.

\section{Conclusion}\label{sec:conclusion}

In this paper, we studied discounted control of asymmetric multi-robot multi-queue systems with switching delays. Exploiting the exhaustive-service structure of the problem, we developed a compact exhaustive-assignment actor-critic policy in which learning is restricted to the reassignment of idle robots, while robots at nonempty queues remain committed to service. This structured design matches the queueing dynamics and yields a more efficient policy class than an unconstrained joint-action representation.

Numerical results across multiple asymmetric arrival-rate profiles and system sizes showed that EA-AC consistently improves upon the ESL baseline in both discounted holding cost and mean queue length, providing empirical evidence that ESL is suboptimal once the symmetry conditions are removed. On instances where optimal benchmarks are available, EA-AC achieves near-optimal performance. The same policy architecture also remains effective across a broad range of problem sizes, indicating favorable practical scalability. Future work will study stronger analytical guarantees, broader generalization across system families, and extensions to weighted-cost formulations.

\section{Appendix}

\subsection{Network Architecture Details}\label{app:network}

Table~\ref{tab:network_arch} summarizes the main architectural hyperparameters of the proposed actor and critic networks used in all experiments. The actor and critic share the same overall hidden width, while using separate encoders for queue and robot features. The actor additionally applies feasibility masks and sequential reservation masks to enforce the assignment constraints described in Section~\ref{sec:policy_network}.

\begin{table*}[t]
\centering
\caption{Architecture summary of the proposed EA-AC policy and value networks.}
\label{tab:network_arch}
\small
\begin{tabular}{l l}
\hline
Robot location embedding dimension & 16 \\
Actor hidden width $d$ & 128 \\
Critic hidden width $d$ & 128 \\
Actor queue encoder & 2-layer MLP \\
Actor robot encoder & 2-layer MLP \\
Actor assignment scorer & scaled dot-product $+\;$queue bias \\
Actor masking & occupancy and reservation masks \\
Critic queue encoder & 2-layer MLP \\
Critic robot encoder & 2-layer MLP \\
Critic pooling & mean pooling over queue and robot tokens \\
Critic value head & MLP $(2d+4,\,128,\,1)$ \\
Activation & ReLU \\
Input preprocessing & normalized queue-length and arrival-rate inputs \\
\hline
\end{tabular}
\end{table*}

\subsection{Additional Scaling Results}\label{app:additional_results}

\begin{figure*}
    \centering
    \includegraphics[width=1\linewidth]{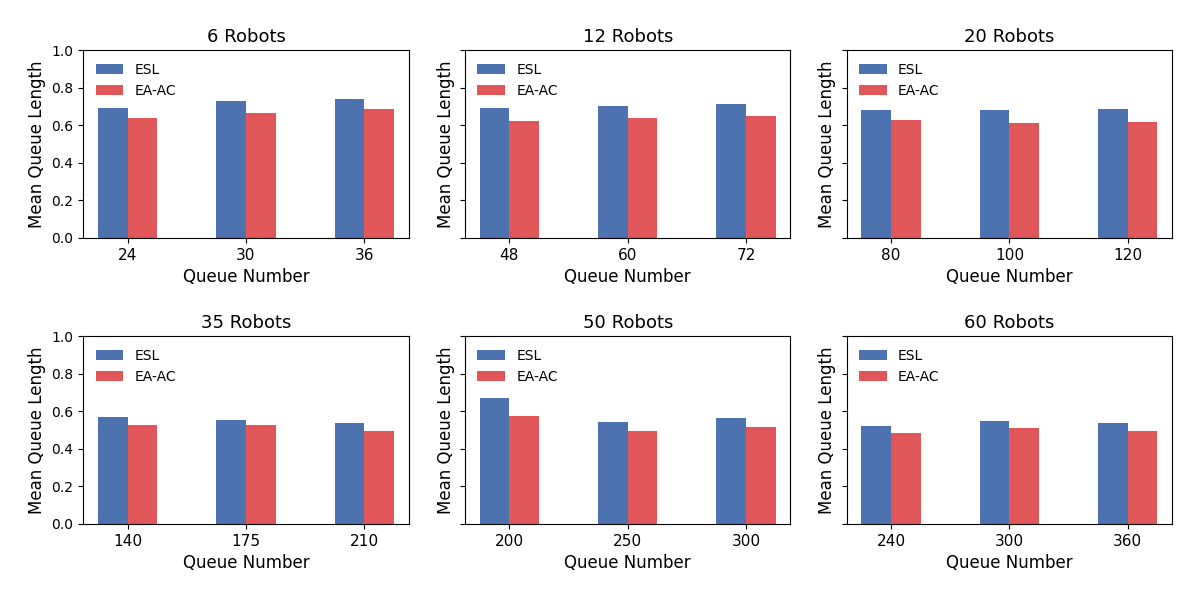}
    \caption{Supplementary comparison of mean queue length for EA-AC and ESL across multiple robot--queue configurations. Each panel corresponds to a fixed number of robots, and the three bars within each panel show performance at different queue counts.}
    \label{fig:additional_results}
\end{figure*}

Figure~\ref{fig:additional_results} provides an additional comparison of mean queue length between EA-AC and ESL across several robot--queue configurations. These results are supplementary to the main performance comparisons reported in Section~\ref{sec:experiments} and are included to illustrate the same qualitative trend over a broader set of system sizes.

\bibliographystyle{IEEEtran}
\bibliography{source}

\end{document}